\documentclass{article}

\usepackage{arxiv}

\usepackage[utf8]{inputenc} 
\usepackage[T1]{fontenc}    
\usepackage{hyperref}       
\usepackage{url}            
\usepackage{booktabs}       
\usepackage{amsfonts}       
\usepackage{nicefrac}       
\usepackage{microtype}      
\usepackage{lipsum}		
\usepackage{graphicx}
\usepackage{natbib}
\usepackage{amsmath}
\usepackage{doi}
\usepackage{float}
\usepackage{natbib}

\title{Welfare Modeling with AI as Economic Agents: A Game-Theoretic and Behavioral Approach}

\author{{\hspace{1mm}Sheyan Lalmohammed} \\
	Unniversity of Pennsylvania\\
	Philadelphia, PA \\
	\texttt{sheyan@wharton.upenn.edu} \\
}

\date{}

\renewcommand{\headeright}{}
\renewcommand{\undertitle}{}
\renewcommand{\shorttitle}{Welfare Model with AI as Economic Agents}

\begin{document}
\maketitle

\begin{abstract}
The integration of artificial intelligence (AI) into economic systems represents a transformative shift in decision-making frameworks, introducing novel dynamics between human and AI agents. This paper proposes a welfare model that incorporates both game-theoretic and behavioral dimensions to optimize interactions within human-AI ecosystems. By leveraging agent-based modeling (ABM), we simulate these interactions, accounting for trust evolution, perceived risks, and cognitive costs. The framework redefines welfare as the aggregate utility of interactions, adjusted for collaboration synergies, efficiency penalties, and equity considerations. Dynamic trust is modeled using Bayesian updating mechanisms, while synergies between agents are quantified through a collaboration index rooted in cooperative game theory. Results reveal that trust-building and skill development are pivotal to maximizing welfare, while sensitivity analyses highlight the trade-offs between AI complexity, equity, and efficiency. This research provides actionable insights for policymakers and system designers, emphasizing the importance of equitable AI adoption and fostering sustainable human-AI collaborations.
\end{abstract}

\section{Introduction}

The introduction of artificial intelligence (AI) systems into economic and social ecosystems traditionally governed under human actions represents one of the most pivotal changes in recent years \citep{agrawal2018prediction}. These systems have rapidly evolved from performing isolated tasks in a narrowly defined way to functioning as collaborative agents capable of augmenting and extending human decisions, automating advanced processes, and generating unforeseen insights \citep{immorlica2024generativeaieconomicagents, li2024econagentlargelanguagemodelempowered}. In domains such as healthcare, supply chain optimization, financial forecasting, and public policy, AI systems, particularly generative AI systems, are becoming pivotal components of decision frameworks \citep{mao2024alympicsllmagentsmeet}. However, the overall transformative process is accompanied by significant challenges, especially in understanding and optimizing the interactions between humans and AI agents \citep{johnson2023evidencebehaviorconsistentselfinterest}. The key question becomes how can we model and measure the optimal behavior between AI agents and humans, who serve a common purpose together to become more efficient economically than either could do alone. The interactions which serve to answer such a question are shaped by unique dynamics, such as trust \citep{mehrotra2023systematicreviewfosteringappropriate}, perceived risk \citep{eec14168-5714-3ca8-b073-d038266f2734}, cognitive costs \citep{Sweller2011}, and emergent synergies \citep{Brusatin_2024}, which classical economic frameworks are ill-equipped to address.

Economic welfare, traditionally narrowly defined as the aggregate utility of individuals within an economic system \citep{samuelson1947foundations}, generally serves as a measure for evaluating societal outcomes after making economic decisions. Welfare functions generally are rooted in utilitarian systems in which generating the most overall utility of all agents in a system is seen as being economically more valuable than a specific subset of individuals in the system receiving a higher utility \citep{a5d507a6-ffc3-3033-b7e2-6e5ae70bb82b}. Classical models of such welfare functions assume homogeneity of the agents being modeled, symmetry in interactions, and stable environmental conditions which allow for simplified aggregation of individual utilities \citep{Steyvers2022}. However, it is clear that the integration of AI agents into human-centric systems, particularly economic systems, is heavily disrupting the generalizations of these assumptions, requiring a rethinking of how the interactions of these agents with human agents can change the way welfare is defined, measured, and optimized \citep{Lindgren16122024}.

In human-AI ecosystems, agent-heterogeneity is central. Humans and AI agents possess fundamentally different capacities in an economic environment \citep{wang2024mutualtheorymindhumanai}. Humans can contribute oversight, judgment, and ethical considerations that AI agents simply lack while AI agents can generally outperform humans in data-processing, pattern recognition, and task execution \citep{dwarakanath2024abideseconomistagentbasedsimulationeconomic}. Furthermore, the interactions between humans and AI agents are asymmetric. A human acts as an approver, evaluator, and user of AI-generated outputs, while AI systems rely on human validation to adapt and improve their performance \citep{dekarske2024dynamichumantrustmodeling}. Part of the interactions between humans and AI are also mediated by behavioral biases previously shown to influence cognitive decision-making, particularly in an economic environment \citep{e6085b40-4f91-3b97-9a48-e38918374ebc}. Humans are prone to show rationality, trust dynamics, and risk aversion which shape their willingness to engage with AI systems \citep{eec14168-5714-3ca8-b073-d038266f2734}. These behaviors are not static, and parameters for decision-making such as trust, AI-generated content reliability, and perceived risk of using AI evolve based on prior interactions and accumulated experience \citep{Steyvers2022, mehrotra2023systematicreviewfosteringappropriate}.

The aforementioned complexities necessitate a paradigm shift in how welfare is modeled for human-AI interaction systems. The goal of such a model would be to optimize the total welfare based on collaborative synergies so that the joint production of human and AI agents exceeds the individual contributions of any single type of agent in the economic environment alone \citep{agrawal2018prediction}. This paper proposes a novel framework to base such a relationship between human and AI agents in an economic setting as to maximize the welfare achieved with the relationship between the two agents. At its core, the framework integrates utility functions meant to theoretically define the utility derived by humans from AI actions and content, adjusted for risk aversion, trust, and costs of use \citep{samuelson1947foundations, Brusatin_2024}. The framework includes collaboration synergies as a core component of measuring additional value (surplus) generated when humans and AI agents collaborate effectively \citep{article, Lindgren16122024}. It also emphasizes the use of inefficiency penalties in the calculation of total welfare in a controlled economic ecosystem, accounting for costs such as cognitive effort and time \citep{Sweller2011}. Finally, the framework considers equity distribution, using penalties to control the variance in utility distribution among human agents to ensure fairness \citep{e6085b40-4f91-3b97-9a48-e38918374ebc}.

\subsection{Theoretical Underpinnings}

The framework builds on classical economic principles while extending them to address the unique characteristics of human-AI systems. Firstly, the framework includes components of behavioral economics, including trust dynamics, cognitive biases, and loss aversion to model realistic decision making by humans with their AI counterparts. Second, there is the inclusion of cooperative game theory which focuses on the development of a collaboration index that prioritizes joint efforts over individual payoffs. Finally, dynamic decision theory is used with trust modeled as a Bayesian updating process, capturing the evolution in the confidence humans will place in AI decision making and generated content over time.  

\subsection{Methodological Approach}

To apply the framework, we develop an agent-based model (ABM) which simulates the interactions between humans and AI agents under varying conditions. The ABM provides a computational platform to quantify the welfare outcomes observed under varying conditions of parameters (trust, information reliability, etc.). These simulations also allow for examination of disparities in utility distribution and the effects of interventions aimed at reducing inequities. Through the use of advanced simulation techniques over multiple discretely defined time periods, sensitivity analyses is enabled, revealing how changes in key parameters to the human-AI interaction affect welfare. 

\subsection{Contributions and Implications}

This paper makes several contributions to the field of economics and decision theory in the context of non-homogeneous agents acting with differing resources toward a common objective under constrained resources. First, it proposes a unique unified welfare framework which combines aspects of traditional welfare concepts with behavioral, operational, and collaborative dimensions unique to human-AI systems. It also integrates dynamic trust modeling into human-AI systems through a Bayesian updating mechanism, providing possible evidence of a modeling tactic for a key aspect of engagement. Lastly, it formalizes the form of a collaboration index for human-AI actions. 

The application of the results, frameworks, and questions raised by this paper are meant to inform the system design of AI-agents in specific-purpose economic environments so that they can be aligned with human preferences. This is also useful for policymakers to determine whether there is need for intervention to ensure equitable AI adoption in specific economic ecosystems. 

\section{Theoretical Framework}

\subsection{Redefining Welfare}

We define total welfare ($W$) as:
$$
W = \underbrace{\sum_{h=1}^{H} \sum_{a=1}^{A} A_h^a \cdot U_h^a}_{\text{Utility from Interactions}} + \underbrace{\phi \cdot \text{Collaboration Index}}_{\text{Productivity Gains}} - \underbrace{\psi \cdot \text{Total Resources Consumed}}_{\text{Efficiency Penalty}} - \underbrace{\alpha \cdot Var(U_h)}_{\text{Equity Penalty}}.
$$

Each term addresses a specific dimension of welfare in human-AI systems. Traditional welfare functions are rooted in utilitarian approaches, aggregating individual utilities \citep{samuelson1947foundations}. However, integrating AI agents into human-centric economic systems necessitates new dimensions, including collaboration synergies and penalties for inefficiencies and inequities \citep{immorlica2024generativeaieconomicagents, Brusatin_2024}.

\begin{itemize}
    \item $H$: Total number of human agents in the system.
    \item $A$: Total number of AI agents in the system.
    \item $A_h^a \in \{0, 1\}$: Approval indicator for whether human $h$ approves the output of AI agent $a$.
    \item $U_h^a$: Utility derived by human $h$ from interacting with AI agent $a$.
    \item $\phi$: Weighting factor for the collaboration index.
    \item $\psi$: Weighting factor for efficiency penalties.
    \item $\alpha$: Weighting factor for equity penalties.
    \item $\text{Variance}(U_h)$: Variance in utility across humans.
\end{itemize}

\subsection{Utility from Interactions}

The utility derived by human $h$ from interacting with AI agent $a$ is defined as:
\begin{equation}
U_h^a = T_h \cdot S_a - \lambda_h \cdot R - C_h,
\end{equation}
where:
\begin{itemize}
    \item $T_h$: Trust level of the human in the AI agent. Trust in AI systems evolves based on prior interactions and can be modeled dynamically using Bayesian updating mechanisms \citep{Steyvers2022, dekarske2024dynamichumantrustmodeling}.
    \item $S_a$: Signal strength, reflecting the reliability of the AI output \citep{johnson2023evidencebehaviorconsistentselfinterest}.
    \item $\lambda_h$: Risk aversion coefficient of the human. Human risk perception and loss aversion are influenced by biases, as explained in prospect theory \citep{eec14168-5714-3ca8-b073-d038266f2734}.
    \item $R$: Perceived risk of engaging with the AI \citep{mao2024alympicsllmagentsmeet}.
    \item $C_h$: Cognitive and temporal costs, defined as:
    \begin{equation}
    C_h = \frac{\text{AI Complexity}}{\text{Human Expertise}} + \frac{1}{\text{Available Time}}.
    \end{equation}

    \begin{itemize}
        \item $\text{AI Complexity}$: Difficulty of understanding or using the AI system. Higher complexity increases cognitive load, as articulated in cognitive load theory \citep{Sweller2011}.
        \item $\text{Human Expertise}$: Skill level of the human, reducing perceived complexity \citep{agrawal2018prediction}.
        \item $\text{Available Time}$: Amount of time the human can devote to the interaction.
    \end{itemize}
\end{itemize}

\subsection{The Approval Function}

The approval function models the decision-making process of humans interacting with AI agents, determining whether the output of AI agent $a$ is approved by human $h$:
\begin{equation}
A_h^a = 
\begin{cases} 
1, & \text{with probability } P(A_h^a = 1) = \frac{1}{1 + e^{-F_h}}, \\
0, & \text{otherwise}.
\end{cases}
\end{equation}
Here:
\begin{itemize}
    \item $A_h^a$: Approval indicator, where $A_h^a = 1$ represents approval and $A_h^a = 0$ represents rejection.
    \item $P(A_h^a = 1)$: Probability of approval, modeled using a logistic function \citep{dwarakanath2024abideseconomistagentbasedsimulationeconomic}.
    \item $F_h$: Approval score, encapsulating the human’s evaluation of the interaction.
\end{itemize}

The approval score $F_h$ is defined as:
\begin{equation}
F_h = \underbrace{\Delta U_h - \lambda_h \cdot \text{Risk} - C_h}_{\text{Behavioral and Cost Factors}}
    + \underbrace{\eta \cdot T_h}_{\text{Dynamic Trust}}
    + \underbrace{\phi \cdot \text{NBS}(U_h, U_a)}_{\text{Game-Theoretic Synergy}}
    + \underbrace{\gamma \cdot S_a}_{\text{AI Signal Strength}}.
\end{equation}

Dynamic trust mechanisms and game-theoretic synergies are critical in this framework. Trust evolves over time based on Bayesian principles \citep{Steyvers2022}, while collaborative gains align with cooperative game theory, particularly the Nash Bargaining Solution (NBS) \citep{article, e6085b40-4f91-3b97-9a48-e38918374ebc}.

\subsection{Collaboration Index}

The collaboration index quantifies synergies that arise from collaboration:
\begin{equation}
\text{Collaboration Index} = \phi \cdot \sum_{h=1}^{H} \sum_{a=1}^{A} \max(0, U_h^a - U_h^{\text{independent}} - U_a^{\text{independent}}),
\end{equation}
where collaborative gains are rooted in joint productivity, as discussed in recent explorations of human-AI complementarity \citep{Steyvers2022, mao2024alympicsllmagentsmeet}.

\subsection{Efficiency and Equity Penalties}

Efficiency penalties are designed to capture the cognitive and temporal costs associated with human-AI interactions:
\begin{equation}
\text{Efficiency Penalty} = \psi \cdot \sum_{h=1}^{H} C_h.
\end{equation}
The framework builds on theories of cognitive effort and resource limitations, emphasizing that collaboration must account for human constraints \citep{Sweller2011, agrawal2018prediction}.

Equity penalties ensure fairness by penalizing disparities in outcomes:
\begin{equation}
\text{Equity Penalty} = \alpha \cdot \text{Var}(U_h),
\end{equation}
aligning with economic principles emphasizing the importance of equity in welfare distributions \citep{e6085b40-4f91-3b97-9a48-e38918374ebc, a5d507a6-ffc3-3033-b7e2-6e5ae70bb82b}.

\section{Agent-Based Modeling Framework}

Agent-Based Modeling (ABM) was chosen as the primary framework in which to analyze and test the model presented in the previous section as a proper way of present the human-AI relationship quantitatively. The reason for this is that ABM allows for a granular decomposition of the human-AI economy away from the natural tendency of equilibrium models in classical economics to achieve a steady state under some specified conditions. Essentially, ABM allows the testing of complicated, and possibly unlikely, economic scenarios by changing the characteristics, or parameters, of human and AI agents to capture the complexity of decision-making, trust evolution, and collaborative-synergies over a discretely defined time period which is unspecified. 

\subsection{Initialization of Human and AI agents}

The initialization of both human and AI agents in the simulated economy assigns unique characteristics to each agent, allowing for the reflection of the heterogeneity in the economy. The characteristics are drawn from a specified probability distribution reflecting the assumptions of the expectations of a human-AI relationship not deeply explored in the current literature. The initialization assumes a population of 100 human agents and 500 AI agents, reflecting the imbalance of the two types of agents in real-world economic systems where AI agents will significantly outnumber human users. Note, the standardized use of a uniform distribution for agent parameters reflects the unknown true distribution in the economy.

\subsubsection{Human Agent Initialization}

The following are the conditions, distributions, and assumptions under which human agents are initialized in the simulated economy: 

\begin{itemize}
    \item \textbf{Trust ($T_h$)} is initialized from a uniform distribution \( T_h \sim U(0.5, 1.0) \), capturing moderate baseline trust in AI systems. This range reflects the assumption that initial trust levels are neither excessively high nor low, allowing for dynamic evolution through interactions.
    \item \textbf{Loss Aversion ($\lambda_h$)} is sampled from \( \lambda_h \sim U(0.5, 1.0) \), representing individual sensitivity to perceived risks. This parameter accounts for differences in behavioral responses to uncertainty and risk in economic environments.
    \item \textbf{Available Time} is drawn from \( \text{Available Time} \sim U(5, 20) \), representing the cognitive and temporal constraints faced by human agents. A wider range ensures that the population includes both highly constrained and more flexible agents.
    \item \textbf{Expertise} is sampled from \( \text{Expertise} \sim U(1.0, 2.0) \), quantifying human skill levels that mitigate cognitive costs when interacting with AI systems.
\end{itemize}

\subsubsection{AI Agent Initialization}

The following are the conditions, distributions, and assumptions under which AI agents are initialized in the simulated economy: 

\begin{itemize}
    \item \textbf{Signal Strength ($S_a$)} is initialized from a uniform distribution \( S_a \sim U(0.8, 1.2) \), reflecting the variability in the reliability of AI outputs. This range ensures that some AI agents provide outputs of higher quality than others, a realistic feature of AI systems in practice.
    \item \textbf{Complexity} is drawn from \( \text{Complexity} \sim U(0.8, 1.5) \), representing the difficulty of interpreting AI outputs. Higher complexity imposes greater cognitive costs on human agents, influencing their likelihood of approving AI-generated outputs.
\end{itemize}

\subsection{Assumptions in the Human-AI Economy}

The ABM incorporates several key assumptions to reflect the dynamics of human-AI interactions in an economic context. Firstly, the ABM assumes that the human agents and AI agents take on distinct roles in their relationship. No AI agent can make an economic decision without human approval. The determination of giving that approval come from the approval function. The asymmetry in this relationship follows real rules of current implementations of AI agents in workplaces.

The next key assumption is that trust is modeled as a dynamic variable using Bayesian updating. We assume that a relationship between a human and AI agent develops over time, similar to the relationship between two human agents. This also reflects the iterative learning process inherent to human-AI relationships. Particularly, we place a moderate level on initial trust levels, emphasizing the need for early interactions in the relationship to be more meaningful to the trajectory of trust evolution.

Third, the model assumes that human agents exhibit cognitive and temporal constraints, capturing the results of these constraints through interaction costs. These costs ensure that welfare outcomes are maintained to limitation considering the costs to humans in collaborating or using information from AI agents. The AI agents act independently of one another, allowing for isolation without coordination to the welfare and approvals exhibited in the economy.

Finally, a uniform baseline is assumed for risk for all human agents which is mitigated through individual differences in loss aversion. By allowing risk levels to remain the same in the economy, we assume a simplified version of risk perception but maintain heterogeneity in human decisions.

\subsection{Dynamics of Model}

The ABM approach simulates interactions between human and AI agents over discrete time steps, capturing the iterative relationship and changes to welfare occurring at each distinct time step. During each time step, human agents are randomly paired with AI agents and have to evaluate their outputs. Approval decisions are then made probabilistically using the criteria specified in the theoretical framework. 

The approvals of certain actions and information from AI agents contributes to the dynamic system as they generate feedback. For human agents, approval of actions increases trust through the Bayesian updating mechanism. For AI agents, approval increases signal strength which reflects the ability of AI systems to learn the preferences of output to humans. This positive feedback loop reinforces the relationship between humans and AI over time, theoretically increasing welfare in the system as a whole.

Welfare in the human-AI economy emerges as the results as an aggregate measure of total well-being in the relationships between human and AI agents. It adequately reflects the interplay between costs, utilities, and collaborative synergies.

\section{Results}

The results of the simulation offer valuable insights into the dynamic interactions between human and AI agents and their implications for overall welfare in the modeled economy. By analyzing welfare outcomes, approval rates, and the effects of key parameters through sensitivity analysis, we identify critical factors that shape human-AI collaboration and their broader economic implications.

\subsection{Welfare Trends Over Time}

\begin{figure}[H]
    \centering
    \includegraphics[width=0.7\linewidth]{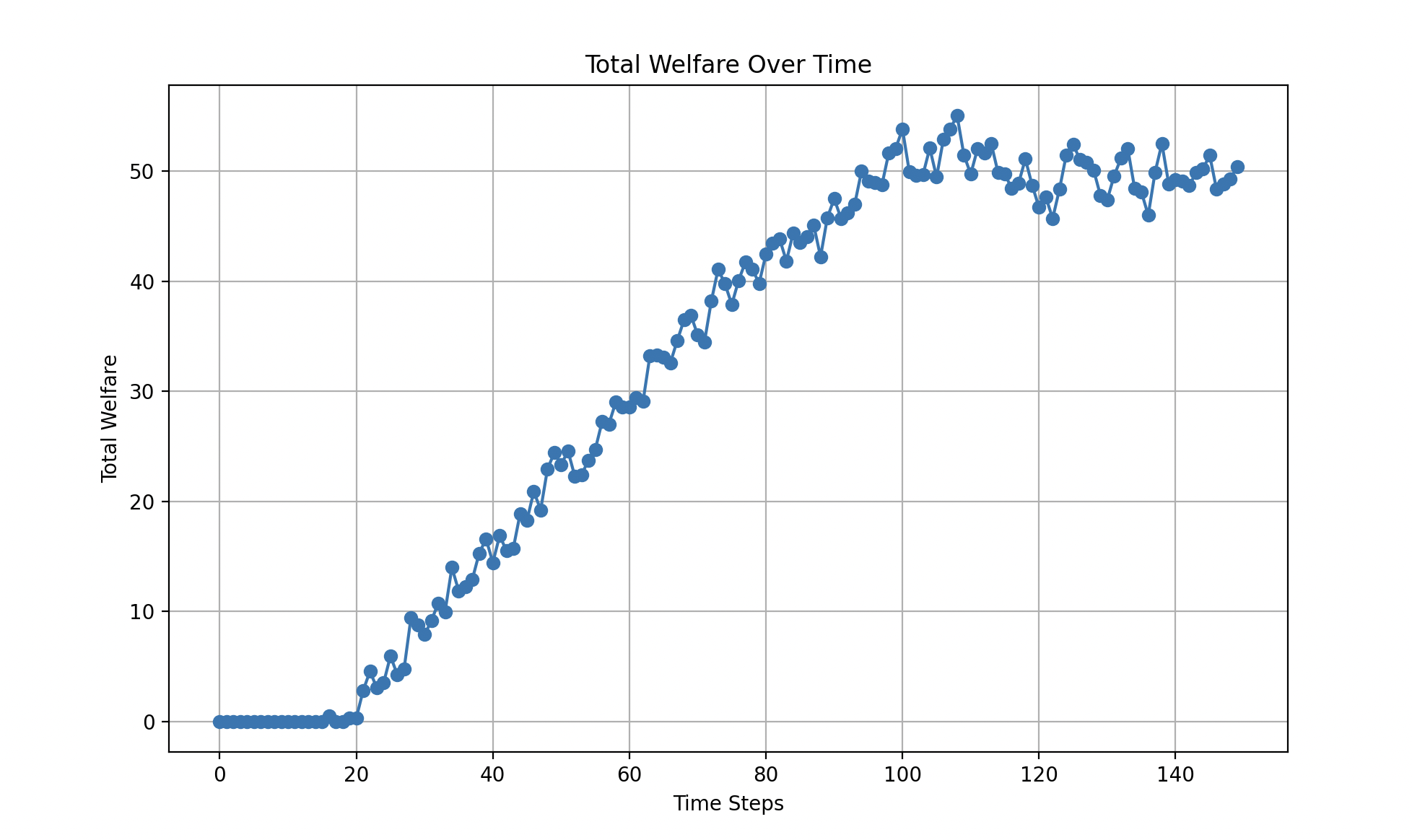}
    \caption{Welfare over 150 Time Steps}
\end{figure}

The evolution of welfare over time, as depicted in Figure 1, demonstrates the gradual improvement in the economic performance of the human-AI system. Welfare begins at a baseline level where interactions are infrequent and characterized by limited trust between agents. Over time, trust dynamics, modeled as a Bayesian updating process, allow human agents to develop confidence in AI outputs. This increasing trust, along with the AI’s adaptive signal improvements driven by human feedback, result in a steady rise in welfare.

The trajectory of welfare speaks to the nature of the environment at each time step. The trajectory is initially steep as human agents overcome barriers to engagement with AI agents. As interactions stabilize and agents adapt to each other, the rate of welfare improvement also begins to stabilize, reflecting the economic system approaching equilibrium. The long-run steady state observed in the simulation is consistent with economic theories of diminishing marginal returns which means as the system approaches the steady state, the incremental gains from further collaboration are reduced.

\subsection{Dynamics of Approval Rates}

\begin{figure}[H]
    \centering
    \includegraphics[width=0.7\linewidth]{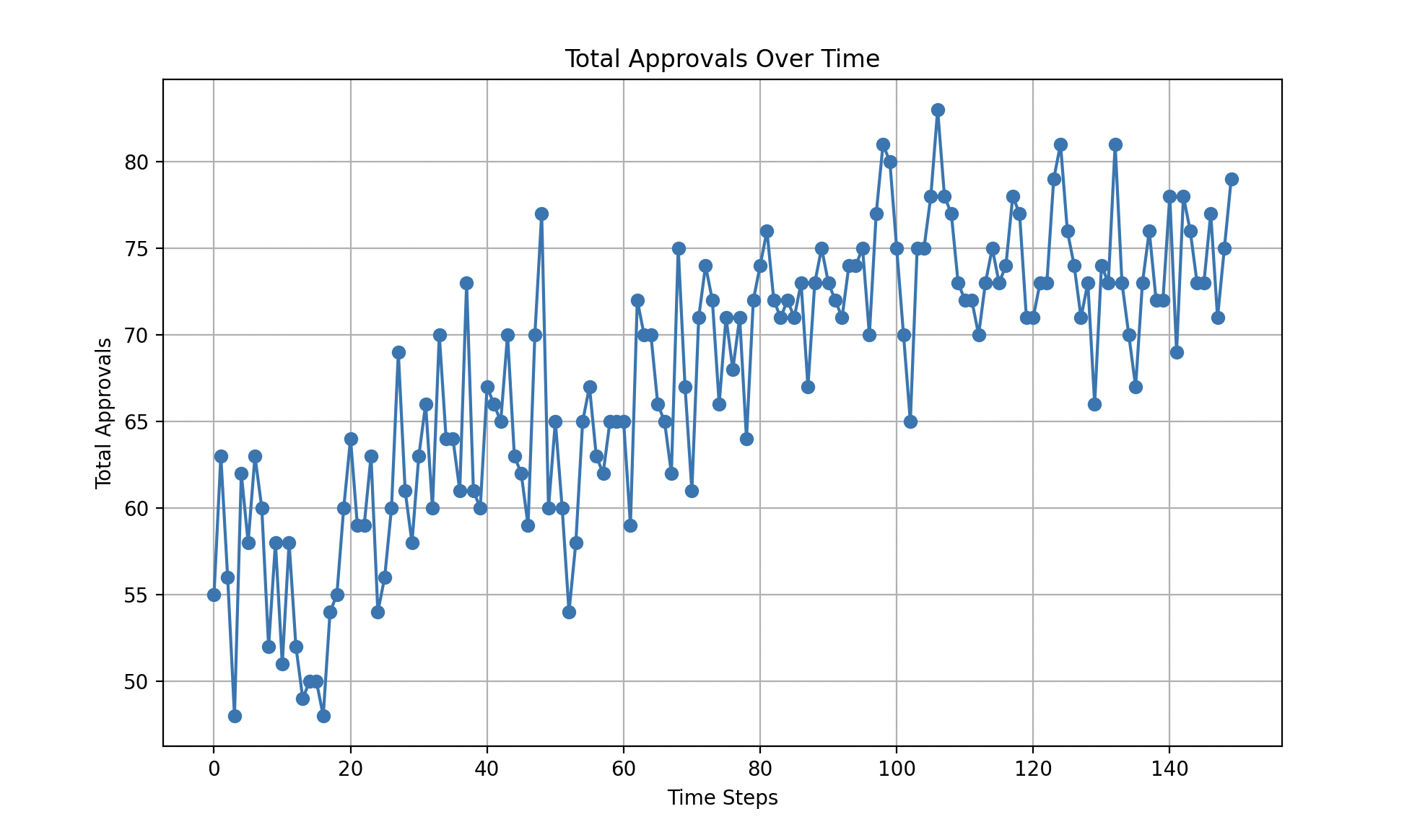}
    \caption{Approvals over 150 Time Steps}
\end{figure}

The progression of approval rates over time, shown in Figure 2, reveals a similar pipeline to what we saw in the welfare trends. Approvals increase gradually as human agents gain familiarity and trust with AI outputs through repeated interactions. Early fluctuations in the approvals highlight the uncertainty inherent in initial interactions, where trust and reliability are still evolving. Over time, this uncertainty diminishes, and approval rates stabilize.

The direct relationship between approvals and welfare underscores the centrality of human validation in determining economic outcomes. Higher approval rates signal not only greater trust but also the effective adaptation of AI agents to human expectations, driving collaborative synergies. The stabilization of approvals further validates the robustness of the proposed trust dynamics and utility framework.

\subsection{Sensitivity to AI Complexity}

\begin{figure}[H]
    \centering
    \includegraphics[width=0.7\linewidth]{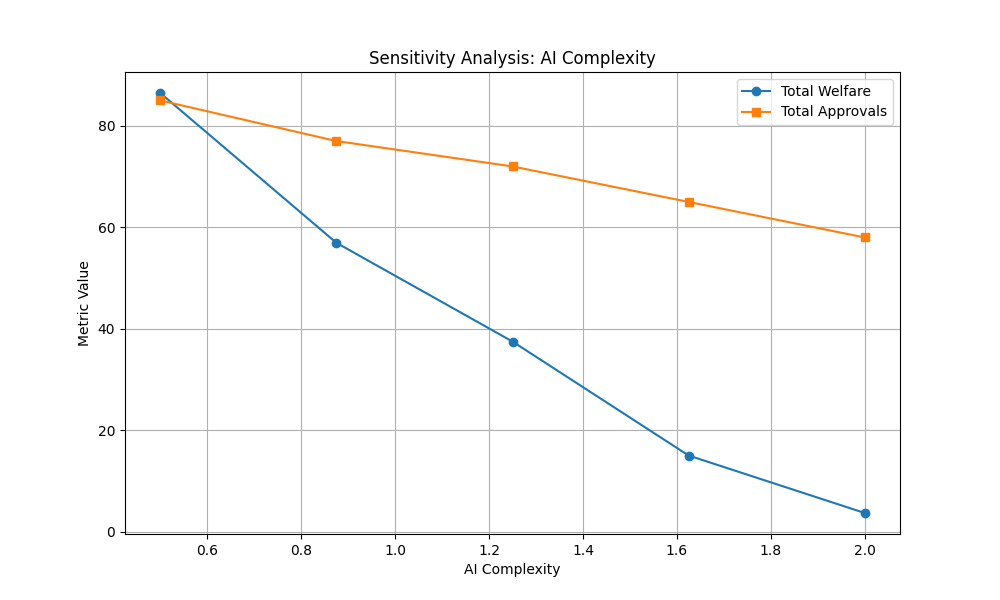}
    \caption{AI Complexity over 150 Time Steps}
\end{figure}

The sensitivity analysis of AI complexity, illustrated in Figure 1, highlights the significant influence of cognitive demands on welfare and approvals. As AI complexity increases, both metrics decline sharply, indicating the adverse effects of high cognitive loads on human agents. The model captures this dynamic through the cost function, where cognitive and temporal burdens reduce utility, leading to fewer approvals and lower welfare.

These results emphasize the importance of human-centered design in AI systems. Simplified and intuitive interfaces, combined with mechanisms to reduce perceived complexity, are essential for fostering productive human-AI collaborations. The findings also suggest a threshold effect: beyond a certain level of complexity, the costs outweigh the benefits, discouraging human engagement and reducing overall system efficiency.

\subsection{Role of Human Expertise}

\begin{figure}[H]
    \centering
    \includegraphics[width=0.7\linewidth]{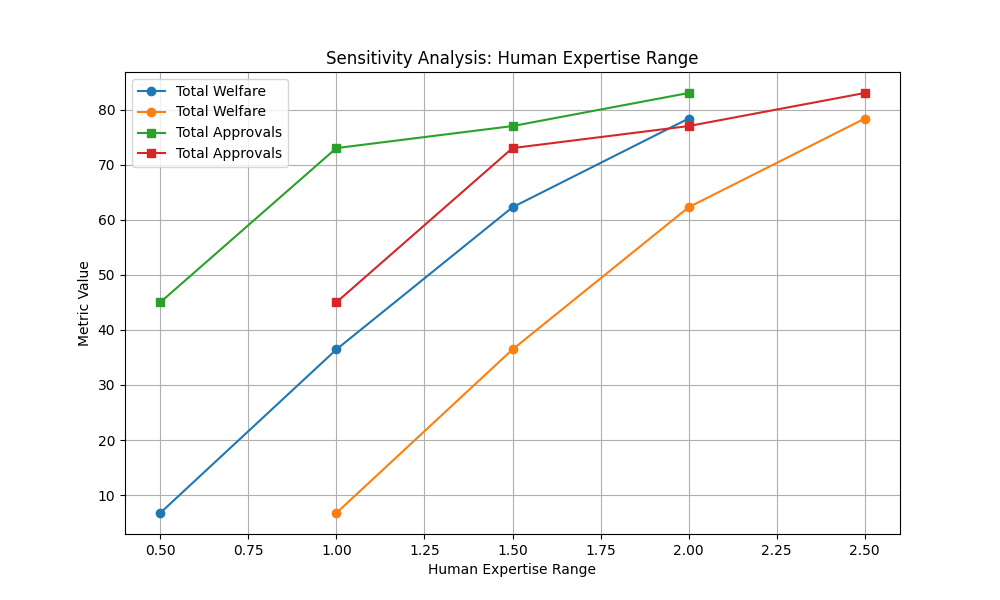}
    \caption{Expertise over 150 Time Steps}
\end{figure}

Human expertise emerges as a critical determinant of welfare and approval outcomes, as shown in Figure 2. The positive relationship between expertise and both metrics underscores the importance of skill development in maximizing the potential of human-AI systems. Higher expertise allows human agents to manage cognitive loads more effectively, derive greater utility from interactions, and approve AI outputs more frequently.

The linear trend in the sensitivity analysis suggests that investments in education and training programs can yield substantial welfare gains. Policies aimed at improving digital literacy and technical skills, particularly among populations with low initial expertise, can enhance the inclusivity and effectiveness of AI adoption. This result aligns with economic theories emphasizing the role of human capital in driving productivity and innovation.

\subsection{Perceived Risk and Its Economic Impact}

\begin{figure}[H]
    \centering
    \includegraphics[width=0.7\linewidth]{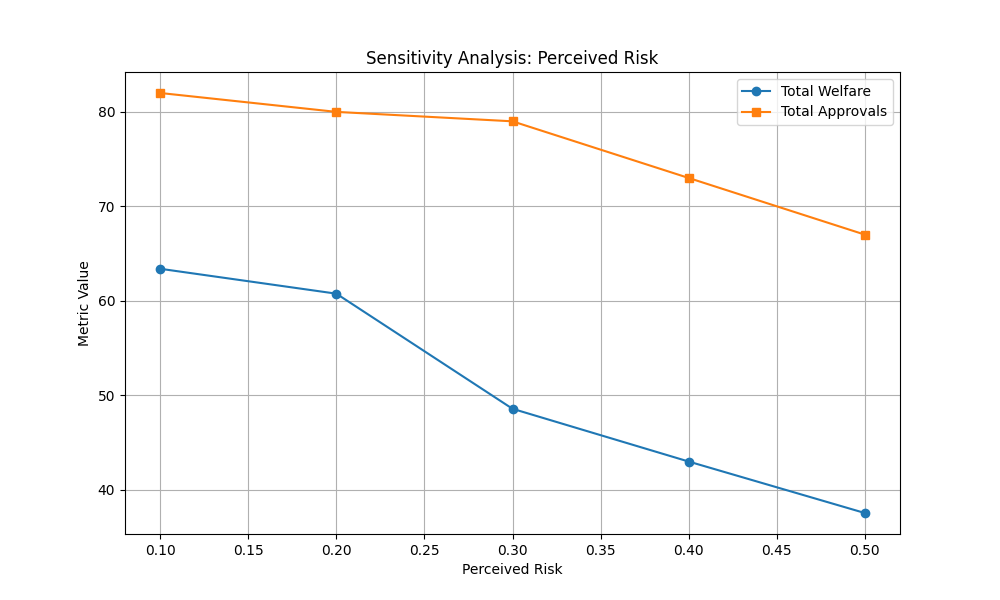}
    \caption{Risk over 150 Time Steps}
\end{figure}

The analysis of perceived risk, presented in Figure 3, reveals its strong negative correlation with welfare and approvals. Higher perceived risks discourage human engagement with AI agents, resulting in fewer interactions and reduced collaborative synergies. The decline in welfare highlights the importance of addressing psychological and informational barriers that limit trust in AI systems.

Interestingly, the results also indicate that trust dynamics partially mitigate the adverse effects of perceived risk. Human agents gradually adjust their risk perceptions based on observed outcomes, suggesting that trust-building mechanisms—such as transparency, accountability, and reliability guarantees—can enhance engagement and mitigate initial hesitations. From a policy perspective, interventions to reduce perceived risks, such as certification standards for AI reliability and ethical safeguards, are essential for fostering equitable and sustainable adoption.

\section{Conclusion and Discussion}

The results of this agent-based model provide robust support for the proposed theoretical framework. The observed trends in welfare, approvals, and sensitivity analyses are consistent with the model’s emphasis on trust dynamics, utility optimization, and collaborative synergies. The findings highlight the multifaceted nature of human-AI interactions, where behavioral, cognitive, and systemic factors interact to shape economic outcomes.

The implications of these results extend beyond the specific context of the simulation. First, they underscore the importance of trust-building as a foundational element of human-AI ecosystems. Dynamic trust mechanisms, such as Bayesian updating, play a critical role in fostering collaboration and driving welfare improvements. Second, the results highlight the need for human-centered design in AI systems, emphasizing simplicity, transparency, and alignment with human capabilities. Finally, the sensitivity analyses provide actionable insights for policymakers and system designers, identifying key levers—such as reducing perceived risks, enhancing human expertise, and managing AI complexity—that can optimize welfare outcomes.

Future research should focus on extending this model to include greater heterogeneity among agents, incorporating diverse economic environments, and exploring the effects of targeted interventions. By addressing these complexities, we can build a more comprehensive understanding of the dynamics driving human-AI collaborations and their implications for economic welfare.

\bibliographystyle{plainnat}
\bibliography{references}

\end{document}